\documentclass[epj]{webofc}
\usepackage[utf8]{inputenc}
\usepackage[varg]{txfonts}   
\usepackage{booktabs}
\usepackage{xcolor}

\newcommand{\be}{\begin{equation}}
\newcommand{\ee}{\end{equation}}
\newcommand{\bea}{\begin{eqnarray}}
\newcommand{\eea}{\end{eqnarray}}

\newcommand{\bie}{\begin{small} \begin{itemize}}

\newcommand{\eie}{\end{itemize} \end{small}}

\definecolor{darkred}{rgb}{0.4,0.0,0.0}
\definecolor{darkgreen}{rgb}{0.0,0.4,0.0}
\definecolor{darkblue}{rgb}{0.0,0.0,0.4}
\usepackage[bookmarks,linktocpage,colorlinks,
    linkcolor = darkred,
    urlcolor  = darkblue,
    citecolor = darkgreen]{hyperref}
\usepackage{subfigure}
\wocname{EPJ Web of Conferences}
\woctitle{Lattice2017}
\begin{document}
%
\selectlanguage{english}
\title{%
Tetraquark resonances computed with static lattice QCD potentials and scattering theory
}
\author{%
\firstname{Pedro} \lastname{Bicudo}\inst{1}\fnsep\thanks{Speaker, \email{bicudo@tecnico.ulisboa.pt} } 
\and
\firstname{Marco} \lastname{Cardoso}\inst{1} 
\and
\firstname{Antje}  \lastname{Peters}\inst{2}
\and
\firstname{Martin}  \lastname{Pflaumer}\inst{2}
\and
\firstname{Marc}  \lastname{Wagner}\inst{2}
}
\institute{%
CFTP, Instituto Superior T\'{e}cnico, Universidade de Lisboa,  Av. Rovisco Pais, 1049-001 Lisboa, Portugal
\and
Goethe-Universit\"at Frankfurt am Main, Institut f\"ur Theoretische Physik, Max-von-Laue-Stra{\ss}e 1, D-60438 Frankfurt am Main, Germany
}
\abstract{%
We study tetraquark resonances with lattice QCD potentials computed for two static quarks and two dynamical quarks, the Born-Oppenheimer approximation and the emergent wave method of scattering theory. As a proof of concept we focus on systems with isospin $I = 0$, but consider different relative angular momenta $l$ of the heavy $b$ quarks. We compute the phase shifts and search for $\mbox{S}$ and $\mbox{T}$ matrix poles in the second Riemann sheet. We predict a new tetraquark resonance for $l = 1$, decaying into two $B$ mesons, with quantum numbers $I(J^P) = 0(1^-)$, mass $m = 10576_{-4}^{+4} \, \textrm{MeV}$ and decay width $\Gamma = 112_{-103}^{+90} \, \textrm{MeV}$.
}
\maketitle


\section{\label{intro}Introduction}

A long standing problem in QCD is to understand exotic hadrons. In this work we specialize in tetraquark systems with two heavy antiquarks $\bar{b} \bar{b}$ and two lighter quarks $q q$, where $q \in \{ u,d,s,c \}$. The existence of bound states has been extensively investigated in the recent past by combining static lattice QCD potentials and the Born-Oppenheimer approximation. A stable $u d \bar{b} \bar{b}$ tetraquark with quantum numbers $I(J^P) = 0(1^+)$ has been predicted \cite{Detmold:2007wk,Wagner:2010ad,Bali:2010xa,
Wagner:2011ev,Bicudo:2012qt,Brown:2012tm,
Bicudo:2015vta,Bicudo:2015kna,
Bicudo:2016ooe} and been confirmed by similar computations using four quarks of finite mass \cite{Francis:2016hui}. Here we extend our investigation by including a new technique from scattering theory, the emergent wave method \cite{Bicudo:2015bra}, and search for possibly existing tetraquark resonances (cf.\ also \cite{Bicudo:2017szl} for more details).


\section{Lattice QCD potentials of two static antiquarks $\bar{Q} \bar{Q}$ in the presence of two lighter quarks $q q$}

In a first step we have computed potentials $V(r)$ of two static antiquarks $\bar{Q} \bar{Q}$ in the presence of two lighter quarks $q q$, where $q \in \{ u,d,s,c \}$, using lattice QCD \cite{Wagner:2010ad,Wagner:2011ev}. There are both attractive and repulsive channels. Most promising with respect to the existence of stable tetraquarks or tetraquark resonances are light quarks $q \in \{ u,d \}$ together with $(I = 0,j = 0)$ or $(I = 1,j = 1)$, where $I$ denotes isospin and $j$ light total angular momentum. The corresponding potentials are not only attractive, but also rather wide and deep \cite{Bicudo:2015vta}.

\begin{figure}[t!] 
\centering
\includegraphics[width=0.7\textwidth]{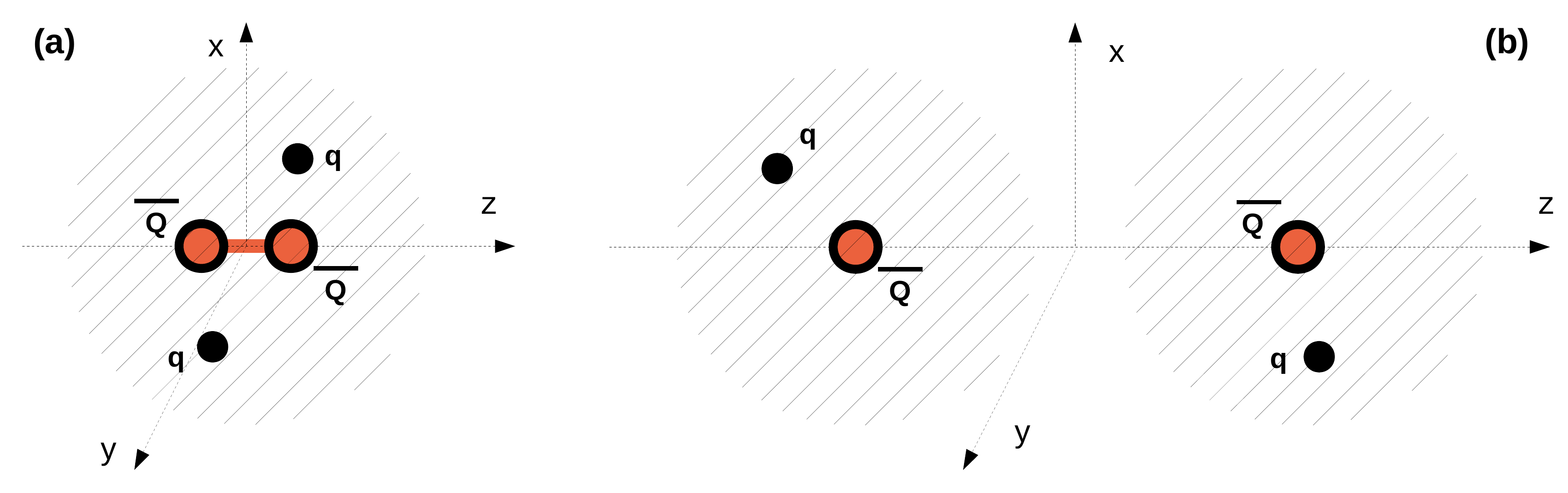}
\caption{(a)~At small separations the static antiquarks $\bar{Q} \bar{Q}$ interact by perturbative one-gluon exchange. (b)~At large separations the light quarks $q q$ screen the interaction and the four quarks form two rather weakly interacting $B$ mesons.}
\label{fig-1}
\end{figure}

\begin{figure}[t!] 
\centering
\includegraphics[width=0.49\textwidth]{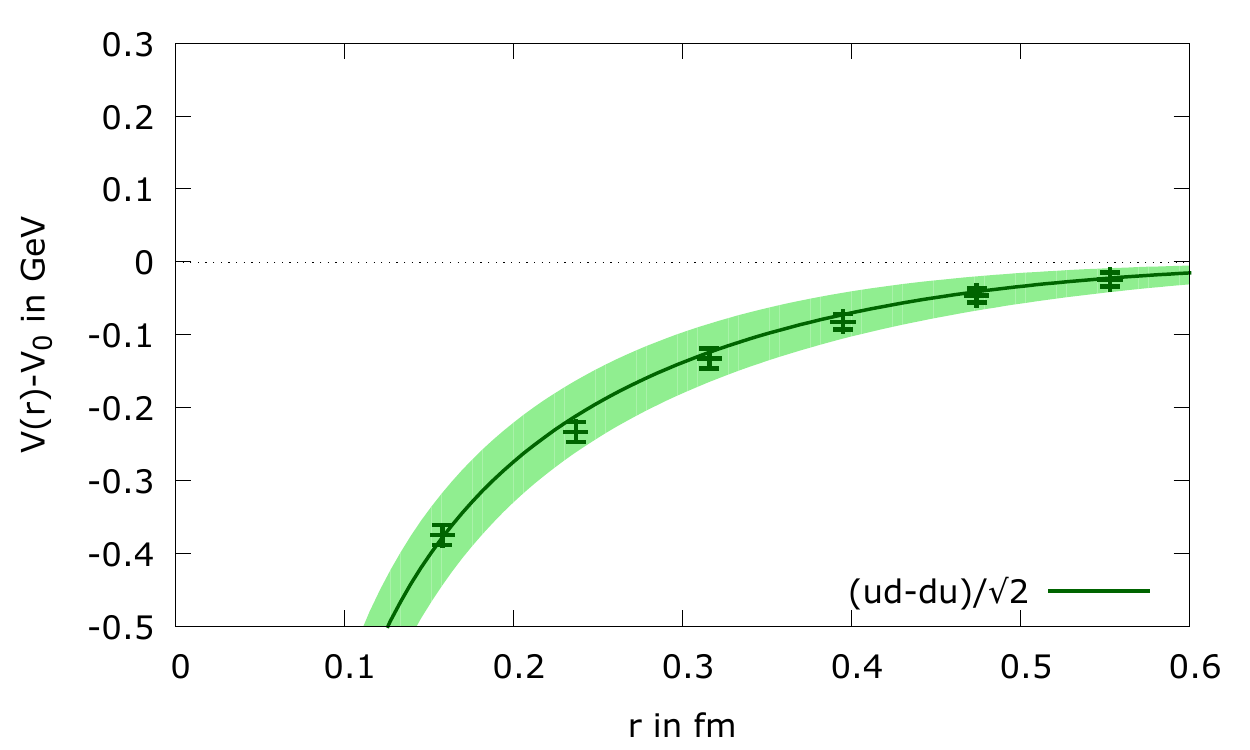}
\includegraphics[width=0.49\textwidth]{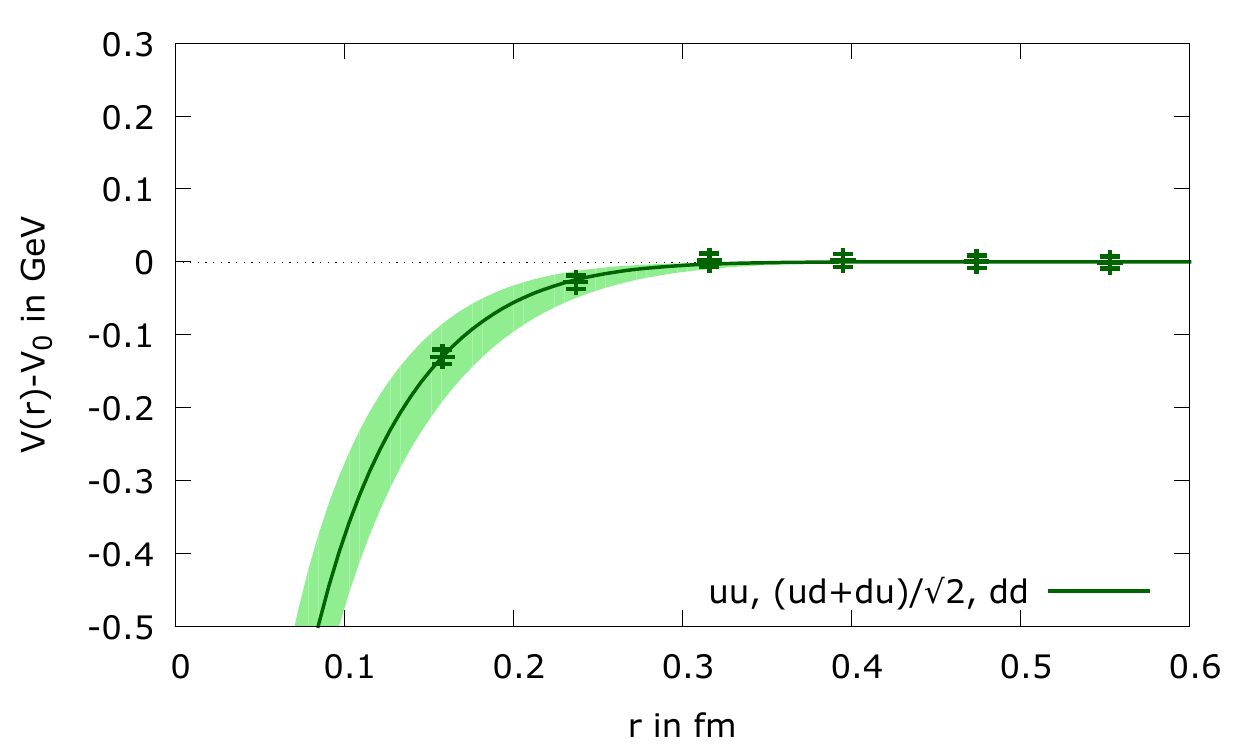}
\caption{(left)~$(I = 0,j = 0)$ potential. (right)~$(I = 1,j = 1)$ potential.}
\label{fig-2}
\end{figure}

The lattice QCD results for the potentials can be parametrized by a screened Coulomb potential,
\begin{equation}
\label{eq:potential}
 V (r) = -\frac{\alpha}{r} e^{-r^2 / d^2}
\end{equation}
inspired by one-gluon exchange at small $\bar{Q} \bar{Q}$ separations $r$ and a screening of the Coulomb potential by the two $B$ mesons at large $r$ (cf.\ Figure~\ref{fig-1}). Clearly, the $(I = 0,j = 0)$ potential is more attractive than the $(I = 1,j = 1)$ potential as can be seen in Figure ~\ref{fig-2}. Numerical results for the parameters $\alpha$ and $d$ are collected in Table~\ref{tab-1}. The potential parametrization is then used in the Schr\"odinger equation for the relative coordinate of the heavy antiquarks $\bar{b} \bar{b} \equiv \bar{Q} \bar{Q}$ to search for either for bound states (cf.\ \cite{Bicudo:2012qt,Bicudo:2015vta,Bicudo:2015kna,
Bicudo:2016ooe}) or for resonances (cf.\ sections ~ \ref{sec:emergent} and ~\ref{sec:results} ).

\section{The emergent wave method
\label{sec:emergent}}

In this section, we summarize the emergent wave method, which is suited to compute phase shifts and resonances. More details can be found in Ref. \cite{Bicudo:2015bra}.

\subsection{Emergent and incident wavefunctions}

We consider the Schr\"odinger equation used for studying bound states:

\begin{equation}
\Big(H_{0} + V(r)\Big) \Psi = E \Psi \ .
\label{eq:schro}
\end{equation}

\begin{table}[t!]
\centering
\begin{tabular}{cc|cc}
\hline
& & & \vspace{-0.25cm} \\
$I$ & $j$ & $\alpha$ & $d$ in $\textrm{fm}$ \\
& & & \vspace{-0.25cm} \\
\hline
& & & \vspace{-0.25cm} \\
$\ 0 \ $ & $\ 0 \ $ & $\ 0.34^{+0.03}_{-0.03} \ $ & $\ 0.45^{+0.12}_{-0.10} \ $ \\
& & & \vspace{-0.25cm} \\
$1$ & $1$ & $0.29^{+0.05}_{-0.06}$ & $0.16^{+0.05}_{-0.02}$\vspace{-0.25cm} \\
& & & \\
\hline
\end{tabular}
\caption{Parameters $\alpha$ and $d$ of the potential of Eq. (\ref{eq:potential}) for two static antiquarks $\bar{Q} \bar{Q}$, in the presence of two light quarks $q q$ with quantum numbers $I$ and $j$.}
\label{tab-1}
\end{table}

The first step in the emergent wave method is to split the wave function of the Schr\"odinger Eq. into two parts,
\begin{equation}
\Psi = \Psi_{0} + X \ ,
\label{eq:sep_psi}
\end{equation}
where $\Psi_{0}$ is the incident wave, a solution of the free Schr\"odinger equation, $
H_{0} \Psi_{0} = E \Psi_{0} $,
and $X$ is the emergent wave. Inserting this in Eq. (\ref{eq:schro}), we obtain:
\begin{equation}
\Big(H_{0} + V(r) - E\Big) X = -V(r) \Psi_{0} \ .
\label{eq:schro_scatter}
\end{equation}

For any energy $E$ we calculate the emergent wave $X$ by providing the corresponding
$\Psi_{0}$ and fixing the appropriate boundary conditions. From the asymptotic behaviour of the emergent wave $X$ we then determine the phase shifts $\delta_l$ , the $\mbox{S}$ matrix and the $\mbox{T}$ matrix. Continuing to complex energies $E \in  \mathbb{C} $ we find the poles of the $\mbox{S}$ matrix and the $\mbox{T}$ matrix in the complex plane. We identify a resonance with a pole of $\mbox{S}$ in the second Riemann sheet at $m - i \Gamma/2$, where $m$ is the mass and $\Gamma$ is the resonance decay width.

\subsection{Partial wave decomposition}

The two heavy antiquarks $\bar{b}\bar{b}$ at zero total angular momentum are described by the Hamiltonian:

\begin{equation}
H = H_0 + V(r) = -\frac{\hbar^{2}}{2 \mu} \triangle + V(r)
\label{EQN005}
\end{equation}
with reduced mass $\mu = M/2$, where $M = 5 \, 280 \, \textrm{MeV}$ is the mass of the $B$ meson from the PDG \cite{Agashe:2014kda}. For simplicity we omit the additive constant $2 M$ in Eq. (\ref{EQN005}), i.e.\ all resulting energy eigenvalues are energy differences with respect to $2 M$.

We consider an incident plane wave $\Psi_{0} = e^{i \mathbf{k} \cdot \mathbf{r}}$, which can be expressed as a sum of spherical waves,
\begin{equation}
\Psi_{0} = e^{i \mathbf{k} \cdot \mathbf{r}} = \sum_{l} (2l+1) i^{l} j_{l}(k r) P_{l}(\hat{\mathbf{k}} \cdot \hat{\mathbf{r}}) \ ,
\label{eq:expansionsphericalbessel}
\end{equation}
where $j_{l}$ are spherical Bessel functions, $P_{l}$ are Legendre polynomials and the relation between energy and momentum is $\hbar k = \sqrt{2 \mu E}$. For a spherically symmetric potential $V(r)$ as in Eq. (\ref{eq:potential}) and an incident wave $\Psi_{0} = e^{i \mathbf{k} \cdot \mathbf{r}}$, the emergent wave $X$ can also be expanded in terms of Legendre polynomials $P_{l}$,
\begin{equation}
X = \sum_{l} (2l+1) i^{l} \frac{\chi_l(r)}{k r} P_{l}(\hat{\mathbf{k}} \cdot \hat{\mathbf{r}}) \ .
\label{eq:001}
\end{equation}
Inserting Eq. (\ref{eq:expansionsphericalbessel}) and Eq. (\ref{eq:001}) into Eq. (\ref{eq:schro_scatter}) leads to a set of ordinary differential equations for $\chi_l$:
\begin{equation}
\hspace{-0.7cm} \bigg(-\frac{\hbar^2}{2 \mu} \frac{d^{2}}{dr^{2}} + \frac{l (l+1)}{2 \mu r^{2}} + V(r) - E\bigg) \chi_l(r)  
 = -V(r) k r j_l(k r) \ .
\label{eq:1cl0:radial}
\end{equation}

\subsection{Solving the differential equations for the emergent wave}

The potentials $V(r)$, Eq. (\ref{eq:potential}), are exponentially screened, i.e.\ $V(r) \approx 0$ for $r \geq R$, where $R \gg d$. For large separations $r \geq R$ the emergent wave is, hence, a superposition of outgoing spherical waves, i.e.\
\begin{equation}
\frac{\chi_l(r)}{k r} = i \, t_l h_l^{(1)}(k r) ,
\label{eq:002}
\end{equation}
where $h_l^{(1)}$ are the spherical Hankel functions of first kind.

Our aim is now to compute the complex prefactors $t_l$, which will eventually lead to the phase shifts. To this end we solve the ordinary differential equation (\ref{eq:1cl0:radial}). The corresponding boundary conditions are the following:
\begin{itemize}
\item At $r = 0$: $\chi_l(r) \propto r^{l+1}$.

\item For $r \geq R$: Eq. (\ref{eq:002}).
\end{itemize}
The boundary condition for $r \geq R$ fixes $t_l$ as a function of $E$.

We solve it numerically, with two different numerical techniques approaches:
\begin{itemize}
\item [(1)]\quad a fine uniform discretization of the interval $[0,R]$, which reduces the differential equation to a large set of linear equations, which can be solved rather efficiently, since the corresponding matrix is tridiagonal; 
\item [(2)]\quad a standard 4-th order Runge-Kutta shooting method.
\end{itemize}

\begin{figure}[t!] 
\centering
\includegraphics[width=0.66\textwidth]{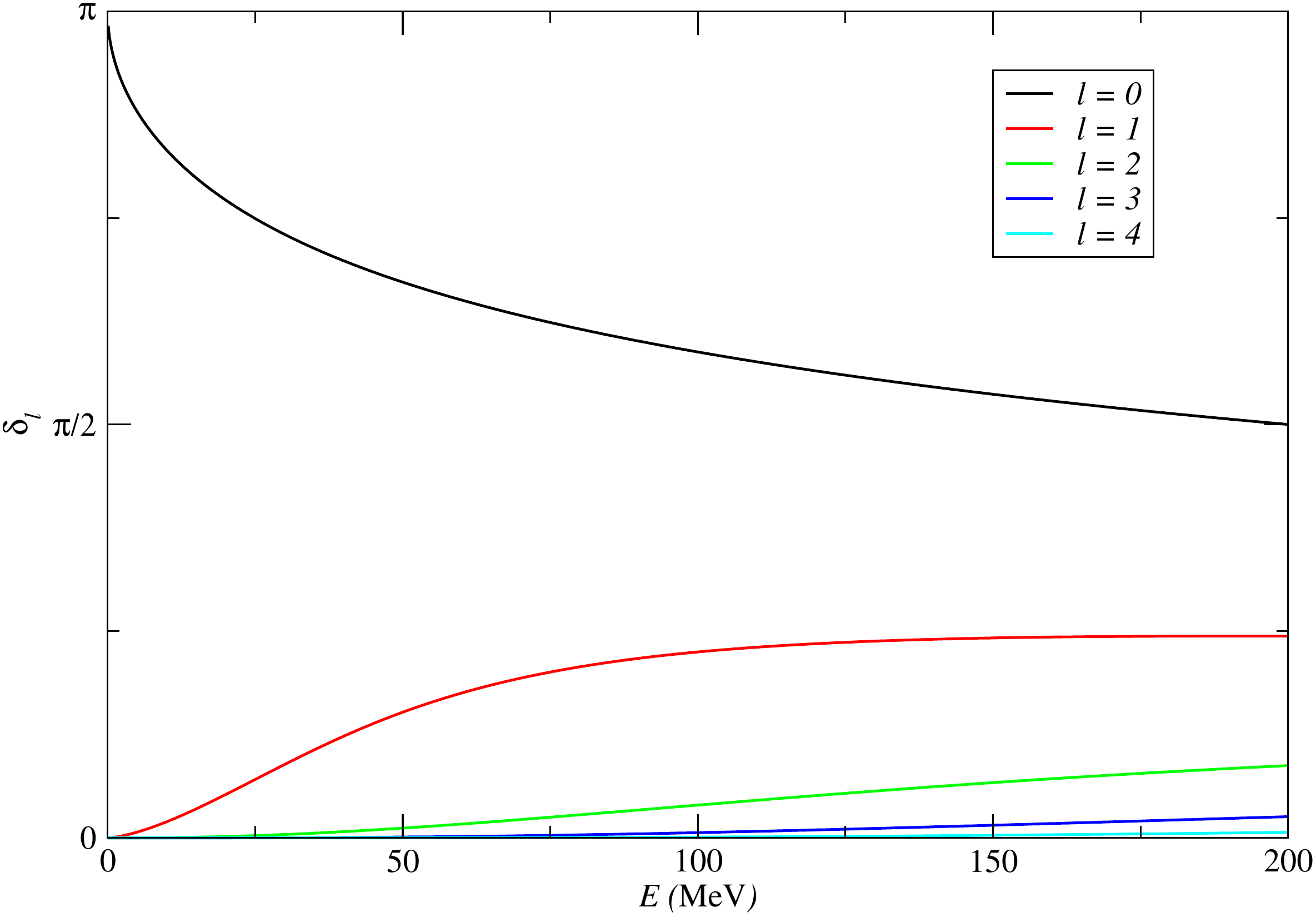}
\caption{
Results for the phase shift $\delta_l$ as a function of the energy $E$ for different angular momenta $l = 0, 1, 2, 3, 4$ for the $(I = 0, j = 0)$ potential ($\alpha = 0.34$, $d = 0.45 \, \textrm{fm}$).}
\label{fig:phaseAngMom}
\end{figure}

\subsection{Phase shifts and $\mathbf{S}$ and $\mathbf{T}$ matrix poles}

The quantity $t_l$ is a $\mbox{T}$ matrix eigenvalue (c.f. a standard textbook on quantum mechanics, e.g. \cite{Merzbacher}). 
For instance, at large distances $r \geq R$, the radial wavefunction is 
\begin{equation}k r [ j_l (kr) + i \, t_l h_l^{(1)}(k r)] = (k r /2) [ h_l ^{(2)}(kr) +e^{2 i \delta_l} h_l^{(1)}(k r)] \ .
\end{equation}
From $t_l$ we can calculate the phase shift $\delta_l$ and also read off the corresponding $\mbox{S}$ matrix eigenvalue $s_l$,
\begin{equation}
s_l \equiv 1 + 2 i t_l = e^{2 i \delta_l} \ .
\label{eq:003}
\end{equation}

Moreover, note that both the $\mbox{S}$ matrix and the $\mbox{T}$ matrix are analytical in the complex plane. They are well-defined for complex energies $E \in  \mathbb{C} $. Thus, our numerical method can as well be applied to solve the differential Eq. (\ref{eq:1cl0:radial}) for complex $E \in  \mathbb{C} $. We find the $\mbox{S}$ and $\mbox{T}$ matrix poles by scanning the complex plane $(\textrm{Re}(E) , \textrm{Im}(E))$ and applying Newton's method to find the roots of $1 / t_l(E)$. The poles of the $\mbox{S}$ and the $\mbox{T}$ matrix correspond to complex energies of resonances. 
Note the resonance poles must be in the second Riemann sheet with a negative imaginary part both for the energy $E$ and the momentum $k$.

\section{Results for phase shifts, $\mathbf{S}$ matrix and $\mathbf{T}$ matrix poles and resonances
\label{sec:results}}

\subsection{Phase shifts}

We first consider the $u d \bar{b} \bar{b}$ potential corresponding to isospin $I = 0$ and light spin $j = 0$ (cf.\ Section \ \ref{sec:emergent}), since this potential is most attractive. We compute $t_l$ and via Eq. (\ref{eq:003}) the phase shift $\delta_l$ for real energy $E$ and angular momenta $l = 0, 1, 2, \ldots$. We do not find a fast increase of the phase shift $\delta_l$ as a function of the energy $E$ which would clearly indicate a resonance (cf.\ Figure\ \ref{fig:phaseAngMom}). 

\begin{figure}[t!] 
\centering
\includegraphics[width=0.66\textwidth]{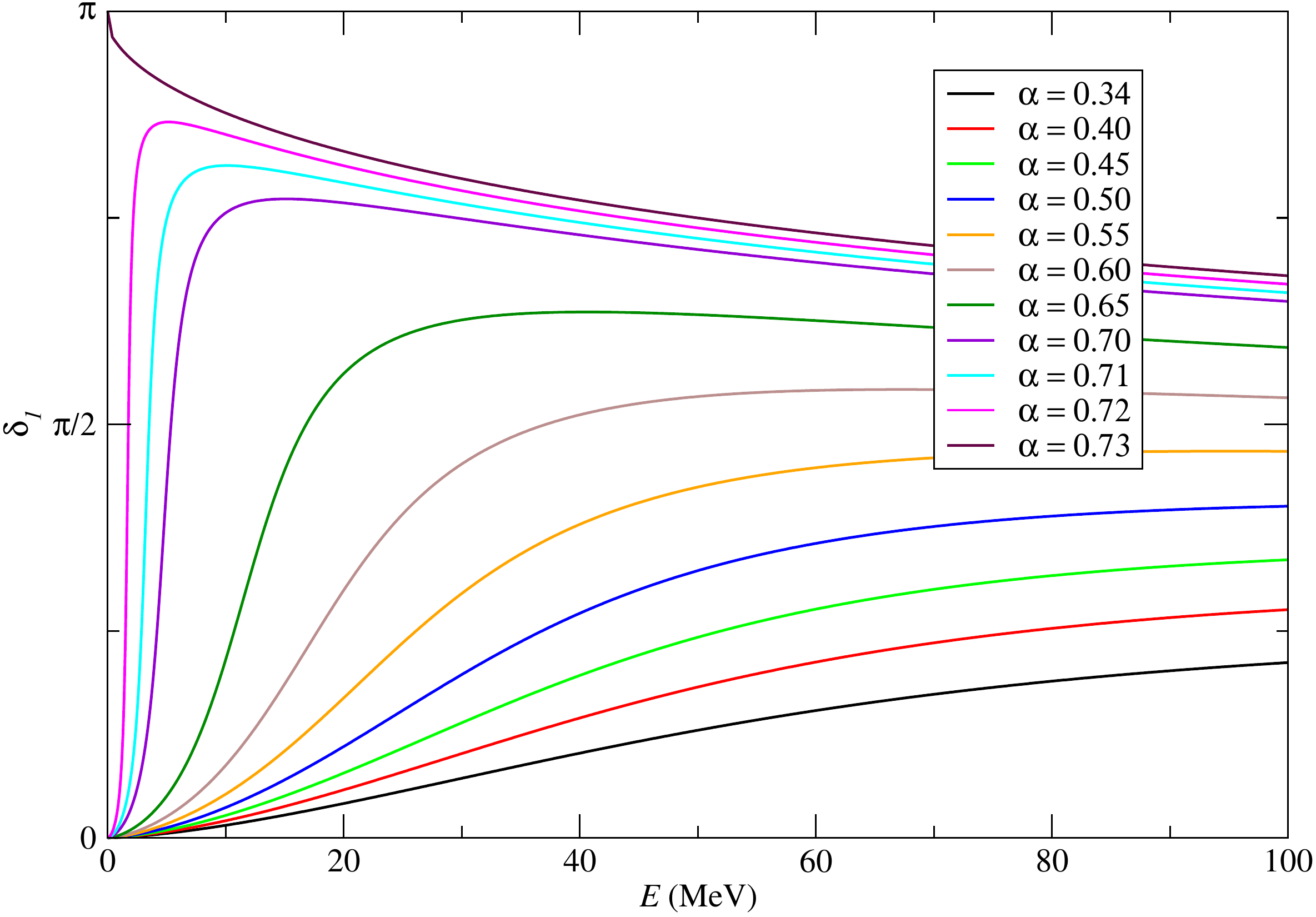}
\caption{
Results for the phase shift $\delta_1$ as a function of the energy $E$ for different parameters $\alpha$ for the $(I = 0, j = 0)$ potential ($d = 0.45 \, \textrm{fm}$).}
\label{fig:delta}
\end{figure}
Thus, we must search more thoroughly for possibly existing resonances. Starting with angular momentum $l = 1$ we first search for clear resonance signals by making the potential more attractive, i.e.\ we increase the parameter $\alpha$. We keep the parameter $d = 0.45 \, \textrm{fm}$ fixed here to preserve the scale of the potential. The corresponding results for the phase shift $\delta_1$ are shown in Figure\ \ref{fig:delta}. Indeed, for $\alpha \approx 0.65$ we find clear resonances. For $\alpha = 0.72$, we find a bound state, since the phase shift $\delta_1$ starts at $\pi$ and decreases monotonically to $0$, when increasing the energy $E$. However, it is not clear from this observation, for which values of $\alpha$ a resonance exists or not.

\subsection{Resonances as poles of the $\mathbf{S}$ and $\mathbf{T}$ matrices } 

\begin{figure}[t!] 
\centering
\includegraphics[width=0.63\columnwidth,trim={0cm 0 0cm 0},clip]{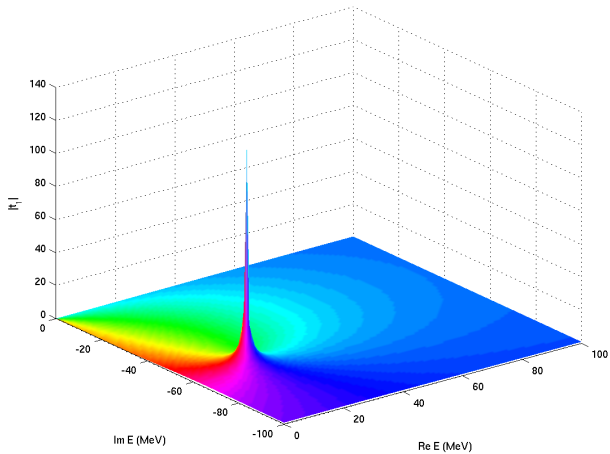}
\caption{
$\mbox{T}$ matrix eigenvalue $t_1$ as a function of the complex energy $E$ . The vertical axis shows the norm $|t_1|$, the colours represent the phase $\textrm{arg}(t_l)$. A pole in the complex plane of $E \in  \mathbb{C} $ is clearly visible. }
\label{fig:peak}
\end{figure}

\begin{figure}[t!] 
\centering
\includegraphics[width=0.60\columnwidth,trim={0 0 0cm 0},clip]{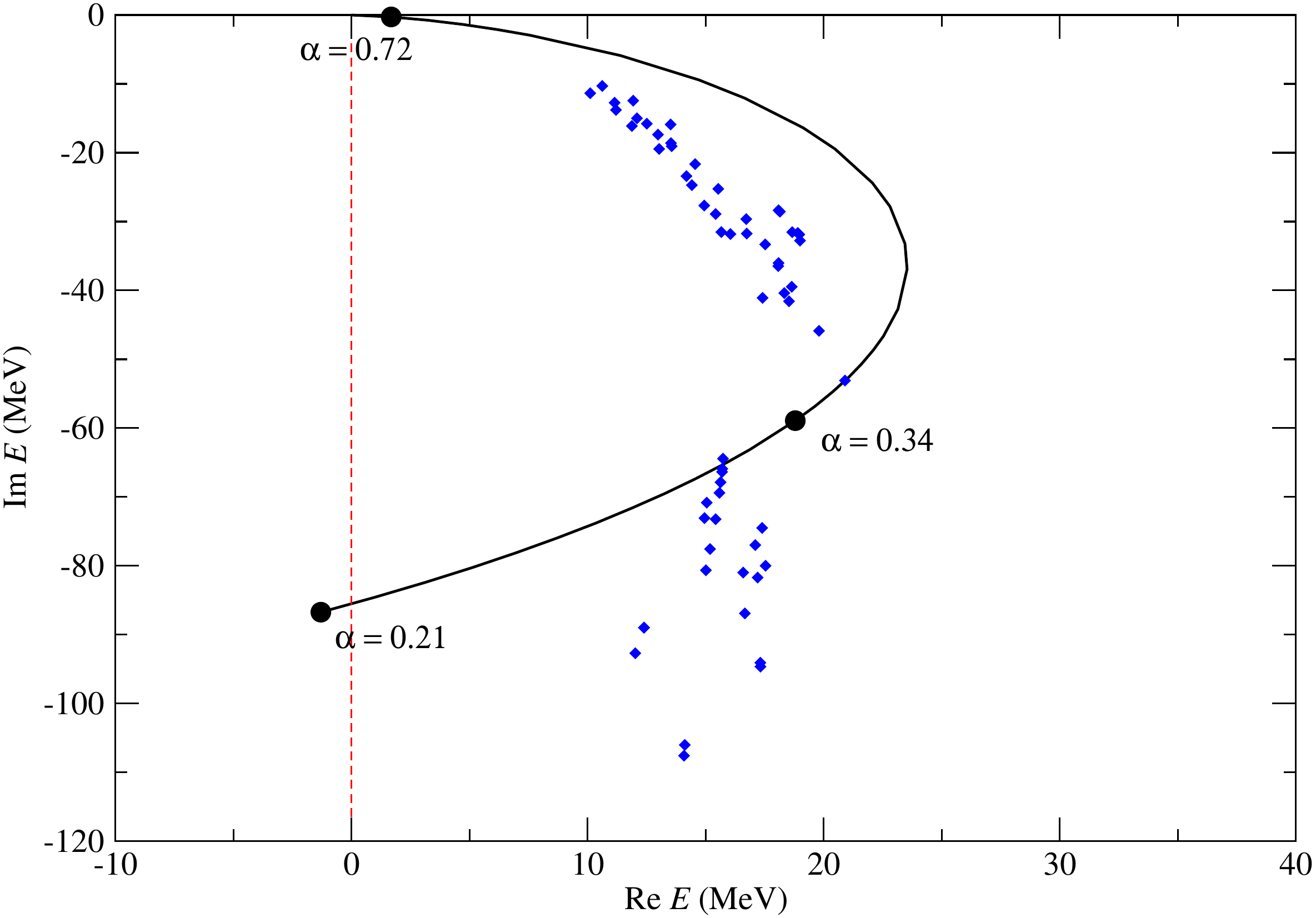}
\caption{
Trajectory of the pole of the eigenvalue $t_1$ of the $\mbox{T}$ matrix  in the complex plane $(\textrm{Re}(E) , \textrm{Im}(E))$, corresponding to a variation of parameter $\alpha$.
We also illustrate with a cloud of diamond points the systematic error 
\cite{Bicudo:2015vta} .}
\label{fig:traj}
\end{figure}

To clearly identify a resonance, we search directly for poles of the $\mbox{T}$ matrix eigenvalues $t_l$. With this technique we clearly find a pole for angular momentum $l = 1$ and physical values of the parameters, $\alpha = 0.34$ and  $d =0.45 \, \textrm{fm}$. We show this pole in Figure\ \ref{fig:peak} by plotting $t_1$ as a function of the complex energy $E$. 
To understand the dependence of the resonance pole on the shape of the potential, we again scan different values of the parameter $\alpha$ and determine each time the pole of the eigenvalue $t_1$ of the $\mbox{T}$ matrix. We show the trajectory of the pole corresponding to a variation of $\alpha$ in the complex plane $(\textrm{Re}(E) , \textrm{Im}(E))$ in Figure\ \ref{fig:traj}. Indeed, starting with $\alpha = 0.21$ we find a pole. This confirms our prediction of a resonance for angular momentum $l = 1$ and physical values of the parameters, $\alpha = 0.34$ and  $d =0.45 \, \textrm{fm}$. In what concerns angular momenta $l \neq 1$, we find no clear signal for a resonance pole (except for the bound state pole for $l = 0$). We also find no poles for any $l$ in the less attractive case of $(I = 1 , j = 1)$.

\subsection{Statistical and systematic error analysis} 

Finally we perform a detailed statistical and systematic error analysis of the pole of $t_1$ and the corresponding values $(\textrm{Re}(E) , \textrm{Im}(E))$ for angular momentum $l = 1$.  We use the same analysis method as for our previous study of the bound state for $l = 0$, cf.\ \cite{Bicudo:2015vta}. To parametrize the lattice QCD data for the potentials, $V^{\textrm{lat}}(r)$, discussed in Section \ref{sec:emergent}, we perform uncorrelated $\chi^2$ minimizing fits with the ansatz of Eq. \eqref{eq:potential}. To this end we minimize the expression
\begin{equation}
\chi^2 = \sum_{r=r_{\textrm{min}},..., r_{\textrm{max}}}\left( \frac{V(r)-V^{\textrm{lat}}(r)}{\Delta V^{\textrm{lat}}(r)}  \right)^2
\label{eq:chisquared}
\end{equation}
with respect to the parameters $\alpha$, $d$ and $V_0$ defined in Eq. (\ref{eq:potential}) and in Refs. \cite{Wagner:2010ad,Wagner:2011ev,Peters:2016wjm}. 
In Eq. (\ref{eq:chisquared}), $\Delta V^{\textrm{lat}}(r)$ denotes the corresponding statistical errors. To quantify systematic errors, we perform a large number of fits. We vary the range of temporal separations $t_{\textrm{min}}\leq t\leq t_{\textrm{max}}$ of the correlation function where $V^{\textrm{lat}}(r)$ is read off as well as the range of spatial $\bar b \bar b$ separations $r_{\textrm{min}}\leq r\leq r_{\textrm{max}}$ considered in the $\chi^2$ minimizing fits to determine the parameters $\alpha$, $d$ and $V_0$.

To also include statistical errors, we compute the jackknife errors of the medians of $\textrm{Re}(E)$ and $\textrm{Im}(E)$ and add them in quadrature to the corresponding systematic uncertainties.

With our combined statistical and systematic error analysis we find a resonance energy $\textrm{Re}(E) = 17^{+4}_{-4} \, \textrm{MeV}$ and a decay width $\Gamma = -2 \textrm{Im}(E) = 112^{+90}_{-103} \, \textrm{MeV}$. Using the Pauli principle and considering the symmetry of the quarks with respect to colour, flavour, spin and their spatial wave function one can determine the quantum numbers of the resonance, which are $I(J^P) = 0(1^-)$. The resonance will decay into two $B$ mesons and, hence, its mass is $m = 2 M + \textrm{Re}(E) = 10 \, 576^{+4}_{-4} \, \textrm{MeV}$.

\section{Summary and outlook}
We utilized lattice QCD potentials computed for two static antiquarks in the presence of two light quarks, the Born-Oppenheimer approximation and the emergent wave method to search for $ud \bar b \bar b$ resonances. First we computed the scattering phase shifts of a  $B B$ meson pair. Then we performed the analytic continuation of the $\mbox{S}$ matrix and the $\mbox{T}$ matrix to the second Riemann sheet, where we have searched for poles $\in \mathbb{C}$. From these results we have predicted a novel resonance with quantum numbers $I(J^P) = 0(1^-)$. Performing a careful statistical and systematic error analysis, we found a resonance mass $m = 10 \, 576^{+4}_{-4} \, \textrm{MeV}$ and a decay width $ \Gamma = 112^{+90}_{-103} \, \textrm{MeV}$. For more details, please see our recent publication \cite{Bicudo:2017szl}.

In the future we plan to address the experimentally observed quarkonia tetraquarks, including $b \bar b$ or $c \bar c$ heavy quarks \cite{Peters:2016wjm}, with our method.


\subsection*{Acknowledgments}

We acknowledge useful conversations with K.~Cichy.

P.B.\ acknowledges the support of CFTP (grant FCT UID/FIS/00777/2013) and is thankful for hospitality at the Institute of Theoretical Physics of Johann Wolfgang Goethe-University Frankfurt am Main. M.C.\ acknowledges the support of CFTP and the FCT contract SFRH/BPD/73140/2010. M.W.\ acknowledges support by the Emmy Noether Programme of the DFG (German Research Foundation), grant WA 3000/1-1.

This work was supported in part by the Helmholtz International Center for FAIR within the framework of the LOEWE program launched by the State of Hesse.

Calculations on the LOEWE-CSC and on the on the FUCHS-CSC high-performance computer of the Frankfurt University were conducted for this research. We would like to thank HPC-Hessen, funded by the State Ministry of Higher Education, Research and the Arts, for programming advice.


\bibliography{literature}

\end{document}